%% file: main.tex
\documentclass[a4paper,conference]{IEEEtran}

\usepackage{cite}
\usepackage{graphicx}
\usepackage{amsmath,amssymb}
\usepackage{url}
\usepackage[abbreviations,nonumberlist]{glossaries-extra}
\setabbreviationstyle{long-short}
\input{acronyms}

\title{Measurement-Calibrated Synthetic Pretraining for Rapid Adaptation of UAV Channel Predictors}

\author{
    \IEEEauthorblockN{
        Qiuheng Zhou\IEEEauthorrefmark{1},
        Wei Jiang\IEEEauthorrefmark{1},
        Donglin Wang\IEEEauthorrefmark{2},
        Hans D. Schotten\IEEEauthorrefmark{1}\IEEEauthorrefmark{2}
    }
    \IEEEauthorblockA{
        \IEEEauthorrefmark{1}
        Intelligent Networks, German Research Center for Artificial Intelligence (DFKI GmbH), Kaiserslautern, Germany\\
        Email: \{qiuheng.zhou, wei.jiang, hans\_dieter.schotten\}@dfki.de
    }
    \IEEEauthorblockA{
        \IEEEauthorrefmark{2}
        Institute for Wireless Communication and Navigation, RPTU Kaiserslautern-Landau, Kaiserslautern, Germany\\
        Email: \{donglin.wang, schotten\}@rptu.de
    }
}

\usepackage{eso-pic}
\AddToShipoutPictureBG*{%
  \AtPageUpperLeft{%
    \raisebox{-1.15cm}{\makebox[\paperwidth][c]{\parbox{0.86\paperwidth}{\centering\scriptsize
    This is the author's accepted version of a paper presented at the 2026 IEEE 104th Vehicular Technology Conference (VTC2026-Fall).\\
    \copyright~2026 IEEE. Personal use of this material is permitted. Permission from IEEE must be obtained for all other uses, in any current or future media, including reprinting/republishing this material for advertising or promotional purposes, creating new collective works, for resale or redistribution to servers or lists, or reuse of any copyrighted component of this work in other works.}}}%
  }%
}

\begin{document}
\maketitle

\begin{abstract}
Unmanned aerial vehicle (UAV) links can change faster than route-specific channel data can be collected, so a channel predictor must often be adapted from only a short target-flight calibration segment. This paper studies measurement-calibrated synthetic pretraining as an offline source prior for rapid adaptation of channel state information (CSI) predictors. The synthetic source is calibrated to flight-only power evolution, temporal shape variation, frequency selectivity, and phase disturbance, and is then combined with measured source-flight fine-tuning before target-flight adaptation. Under chronological held-out flight splits, synthetic-to-real (S2R) fine-tuning reduces the measured-data budget needed to reach useful prediction in the low-data regime. For 50\,ms-ahead prediction, it crosses a latest-channel-estimate persistence baseline with 50 measured training windows, roughly 0.5\,s of 10\,ms-spaced supervised windows, whereas training from scratch requires 600 windows. A target-prefix adaptation study further shows that a lightweight gated recurrent unit predictor can reuse the pretrained source prior with one prediction-layer update epoch, positioning measurement-calibrated synthetic pretraining as a reusable prior for scarce target-flight measurements.
\end{abstract}

\begin{IEEEkeywords}
UAV, channel prediction, synthetic data, pretraining, CSI, measurement-calibrated learning
\end{IEEEkeywords}

\section{Introduction}
Future \gls{6g} systems are expected to support mobile, aerial, and mission-dependent links whose radio conditions change with platform motion and deployment context~\cite{jiang2021road6g}. For \gls{uav} communications, this makes channel prediction attractive for proactive scheduling and link adaptation. In practice, however, a predictor often has to be commissioned before a long route-specific measurement campaign is available. Hardware-inclusive \gls{csi} acquisition is expensive, weather- and mission-sensitive, and difficult to repeat under controlled conditions~\cite{mao2024uavsounding,zhou2026uavdataset}. The operational question is not only whether a learned predictor can perform well with abundant measured training data, but how quickly it can adapt after a short measured calibration segment.

This shifts the emphasis from final accuracy with abundant route-specific training to calibration efficiency. A simple persistence predictor can already be strong for slowly varying magnitude \gls{csi}; synthetic data is valuable only if it reduces the measured-data or update budget needed to cross that no-training reference. We therefore focus on the low-data commissioning phase, where a calibrated prior may help before a target-flight model has enough measurements or update time to learn the channel structure directly.

Prior channel-prediction work has shown that recurrent and attention-based models can exploit temporal channel structure when the training and test conditions are sufficiently aligned~\cite{jiang2020fading,jiang2022transformer,stenhammar2024comparison,zhou2024transformer}. The harder case is realistic transfer: generic fading traces may reproduce broad channel trends while missing the power dynamics, phase disturbance, temporal smoothness, and local irregularity created by receiver-side artifacts and measurement conditions~\cite{joo2019vehicular,zhou2026uavdataset}. Such priors matter only when their statistics are tied to measurement-relevant channel behavior and when their value is assessed under limited real-data adaptation.

We study a measured-flight adaptation setting between pure measurement benchmarking and unconstrained synthetic augmentation. Recent work has explored \gls{gan}-based or generative channel modeling for unknown or measured channels~\cite{jiang2022ragan,euchner2024ganmimo}, and more recent channel-extrapolation studies have discussed generative models as a mechanism for richer \gls{csi} synthesis and transfer~\cite{gao2025multidomain}. In contrast, we use a controllable synthetic source whose power evolution, phase disturbance, and temporal-frequency consistency are tied to measured flight statistics.

The evaluation asks whether this calibrated source reduces the measured adaptation budget. A source-pool study first compares \textit{real-only}, \textit{synthetic-only}, and \textit{\gls{s2r} fine-tuning} under strict held-out splits and explicit measured-data budgets. The rapid-adaptation study then treats the trained source models as cached priors and compares \textit{real-only} and \textit{\gls{s2r}} source priors under the same short target-flight update budget.

The contributions are threefold. First, we formulate a measurement-calibrated synthetic pretraining pipeline for realistic \gls{uav} \gls{csi} prediction. Second, we evaluate it as a rapid adaptation problem using chronological held-out flight splits, explicit measured-data budgets, cached source-prior target-prefix adaptation, and a persistence break-even reference. Third, we identify the operating regime in which calibrated pretraining reduces the measured-data and update budget, while separating this effect from zero-real-adaptation transfer and high-data saturation.

\section{System Model}
Our prediction task is built on hardware-inclusive \gls{uav}-to-ground \gls{csi} measurements collected from a realistic dynamic flight workflow introduced in our prior measurement study~\cite{zhou2026uavdataset}. The acquisition pipeline builds on the open-source OpenAirInterface ecosystem~\cite{kaltenberger2019oai}, whose relevance to non-terrestrial experimentation has also been demonstrated in satellite-oriented \gls{5g} \gls{ntn} studies~\cite{volk2024oai5gntngeo}. Each frame is paired with synchronized telemetry, which is used here to select flight-only, uninterrupted intervals. The present study uses the measured channel tensor after frequency-domain band stitching, retaining a 574-dimensional magnitude target per 10\,ms time step.

We adopt a strict train-test separation within a related measured-flight family. The measured calibration pool consists of separate missions with different motion phases, while a disjoint held-out flight is reserved for evaluation. Flight height is held fixed as a control variable so that the comparison is not dominated by path-loss offsets; the model must still transfer across route realizations and speed profiles rather than memorize a single flight trace. In the intended use case, the budget-limited measured pool represents the short calibration data available before a predictor is commissioned for a related \gls{uav} operating condition.

Let $\mathbf{H}_t\in\mathbb{C}^{K}$ denote the measured frequency-domain \gls{csi} vector at time index $t$ after frame synchronization and channel estimation. Band stitching selects the active subcarrier set $\mathcal{K}$ and maps each frame to
\begin{equation}
    \mathbf{x}_t = |\mathbf{H}_t[\mathcal{K}]| \in \mathbb{R}^{F}, \quad F=574 .
\end{equation}
The measured windows also carry a known radio-frame slot identifier. We encode it as a one-hot timing state $\mathbf{q}_t\in\{0,1\}^{Q}$ with $Q=10$, and use $\mathbf{u}_t=[\mathbf{x}_t,\mathbf{q}_t]$ as the conditioned input when slot metadata are present. Slot indices are deterministic radio-frame positions, not future channel observations; using prediction-time slot states therefore does not reveal future \gls{csi}. Synthetic windows without measured slot metadata use zero timing channels. For a lookback length $L$ and prediction horizon $H$, the input and target windows are
\begin{equation}
    \mathbf{U}_t = [\mathbf{u}_{t-L+1},\ldots,\mathbf{u}_t],\quad
    \mathbf{Y}_t = [\mathbf{x}_{t+1},\ldots,\mathbf{x}_{t+H}],
\end{equation}
and the predictor is a parameterized sequence model conditioned on the known prediction-time slot states $\mathbf{Q}^{+}_t=[\mathbf{q}_{t+1},\ldots,\mathbf{q}_{t+H}]$:
\begin{equation}
    f_{\boldsymbol{\theta}}:\mathbb{R}^{L\times(F+Q)}\times\mathbb{R}^{H\times Q}
    \rightarrow \mathbb{R}^{H\times F}.
\end{equation}
For any dataset $\mathcal{D}$ of input-target windows, the training objective is
\begin{equation}
    \mathcal{L}_{\mathcal{D}}(\boldsymbol{\theta})
    = \frac{1}{|\mathcal{D}|}\sum_{(\mathbf{U},\mathbf{Q}^{+},\mathbf{Y})\in\mathcal{D}}
      \|\mathbf{Y}-f_{\boldsymbol{\theta}}(\mathbf{U},\mathbf{Q}^{+})\|_2^2 .
\end{equation}
The reported metric is \gls{nmse} at the requested prediction horizon on the held-out real test set:
\begin{equation}
    \mathrm{NMSE}_{H} =
    10\log_{10}\frac{\sum\|\mathbf{x}_{t+H}-\hat{\mathbf{x}}_{t+H}\|_2^2}
    {\sum\|\mathbf{x}_{t+H}\|_2^2}.
\end{equation}

\section{Rapid-Adaptation Framework}
\subsection{Training Routes and Target-Prefix Adaptation}
Let $\mathcal{D}_{R}^{\mathrm{cal}}$ be the measured calibration pool and $\mathcal{D}_{R}^{\mathrm{te}}$ the held-out real test set. For a calibration budget $c$, the available measured training set is $\mathcal{D}_{R}^{c}\subset\mathcal{D}_{R}^{\mathrm{cal}}$ with $|\mathcal{D}_{R}^{c}|=c$. Let $\mathcal{D}_{S}$ denote the measurement-calibrated synthetic dataset defined below.

For the main source-pool evaluation, the three training strategies differ only in how the final predictor parameters are obtained. Let
$\mathrm{Train}(\boldsymbol{\theta}^{(0)},\mathcal{D})$ denote the early-stopped optimizer output after training on $\mathcal{D}$ from initialization $\boldsymbol{\theta}^{(0)}$, and let $\boldsymbol{\theta}_{0}$ be the random initialization. The resulting parameter choices are
\begin{align}
    \textit{real-only:}\quad
    \boldsymbol{\theta}_{R}^{c}
    &= \mathrm{Train}(\boldsymbol{\theta}_{0},\mathcal{D}_{R}^{c}), \\
    \textit{synthetic-only:}\quad
    \boldsymbol{\theta}_{S}
    &= \mathrm{Train}(\boldsymbol{\theta}_{0},\mathcal{D}_{S}), \\
    \textit{\gls{s2r}:}\quad
    \boldsymbol{\theta}_{S\rightarrow R}^{c}
    &= \mathrm{Train}(\boldsymbol{\theta}_{S},\mathcal{D}_{R}^{c}).
\end{align}
All training strategies are evaluated on $\mathcal{D}_{R}^{\mathrm{te}}$. Thus, \textit{real-only} uses the budget-limited measured set from scratch, \textit{synthetic-only} tests zero-real-adaptation transfer, and \textit{\gls{s2r} fine-tuning} uses the same measured budget as \textit{real-only} but starts from the calibrated synthetic predictor.

For the target-prefix adaptation check, we separate source-side prior construction from target-side updates. Let $\mathcal{D}_{R}^{\mathrm{src}}$ denote the measured source-flight pool, $\mathcal{D}_{R}^{\mathrm{tp},c}$ the first \(c\) windows of the target flight, and $\mathcal{D}_{R}^{\mathrm{suf}}$ the later target-flight suffix. The \textit{real-only} source prior and the \textit{\gls{s2r}} source prior are
\begin{align}
    \boldsymbol{\theta}_{R}^{\mathrm{src}}
    &= \mathrm{Train}(\boldsymbol{\theta}_{0},\mathcal{D}_{R}^{\mathrm{src}}), \\
    \boldsymbol{\theta}_{S\rightarrow R}^{\mathrm{src}}
    &= \mathrm{Train}(\boldsymbol{\theta}_{S},\mathcal{D}_{R}^{\mathrm{src}}).
\end{align}
Both are then adapted on the same target prefix:
\begin{equation}
    \boldsymbol{\theta}^{c,e}_{\mathrm{adapt}}
    = \mathrm{Update}_{e}(\boldsymbol{\theta}^{\mathrm{prior}},\mathcal{D}_{R}^{\mathrm{tp},c}),
\end{equation}
where \(e\) is the number of target-prefix update epochs. We evaluate the adapted model on $\mathcal{D}_{R}^{\mathrm{suf}}$. In the prediction-layer adaptation case, $\mathrm{Update}_{e}(\cdot)$ optimizes only the final prediction layer while freezing the recurrent backbone.

\subsection{Evaluation Protocol}
The core predictor evaluation covers six sequence settings, extending prior channel-prediction benchmarks that compare recurrent and attention-based temporal models~\cite{jiang2020fading,jiang2022transformer,zhou2024transformer}:
\[
\begin{aligned}
(\text{lookback},\text{horizon}) \in \{&(10,1), (10,2), (10,5),\\
&(10,8), (10,15), (20,5)\}.
\end{aligned}
\]
Because frames are spaced by 10\,ms, lookback values of 10 and 20 correspond to 100 and 200\,ms of observed \gls{csi} history, while horizons of 1, 2, 5, 8, and 15 correspond to 10, 20, 50, 80, and 150\,ms-ahead prediction. These settings let us compare both horizon growth at fixed lookback and lookback variation at fixed horizon. For each setting, we evaluate three training strategies:
\textit{real-only}, \textit{synthetic-only}, and \textit{\gls{s2r} fine-tuning}.

We limit the amount of available measured training data to emulate short calibration campaigns. The calibration budgets are
\[
c \in \{50, 100, 300, 600, 1000, 2000, 3000, 4000, 5000\}
\]
sequence windows. With a 10\,ms frame spacing and unit window stride, \(c\) supervised windows correspond to roughly \(0.01c\) seconds of window-start coverage, in addition to the lookback context and future labels required by the first usable window. The 50- and 100-window budgets represent the scarce-data regime, while 2000--5000 windows expose high-data saturation. The measured budgets are applied only to the training split and allocated across the three measured calibration missions. For \textit{synthetic-only}, the same budget axis is used only as a paired comparison index because no measured training samples are consumed.

To avoid overly optimistic evaluation, we use a chronological split protocol. Sequence windows are first ordered in time, then divided into train, validation, and test segments with an explicit boundary gap of $\text{lookback}+\text{horizon}-1$ windows. This removes direct overlap across split boundaries. The primary metric reported throughout the paper is \gls{nmse} at the requested prediction horizon on the held-out real test set, averaged over three random seeds.

In addition to the main source-pool calibration protocol, we run a focused target-prefix adaptation check. There, the measured fine-tuning budget is taken from the first few seconds of the target flight and evaluation is performed on the later target-flight suffix. The comparison starts from two cached offline priors: a \textit{real-only} source prior trained only on measured source-flight windows, and an \textit{\gls{s2r}} prior obtained by synthetic pretraining followed by measured-source fine-tuning. Neither source prior is rebuilt when the target flight arrives; only the target-side prediction layer is updated from the observed prefix. This models a commissioning scenario in which a \gls{uav} collects a short prefix of target-flight \gls{csi} and then applies a lightweight update before subsequent channel prediction.

\begin{table}[t]
    \scriptsize
    \centering
    \caption{Evaluation protocol.}
    \label{tab:protocol}
    \begin{tabular}{p{1.7cm}p{5.35cm}}
        \hline
        Item & Setting \\
        \hline
        Measured calibration pool & Three same-height source missions: 15\,m/s phases 1--2 and 20\,m/s long route. \\
        Held-out test & Disjoint same-height 10\,m/s target flight. \\
        Filtering & 10\,ms \gls{csi}; uninterrupted flight-only segments. \\
        Representation & 574-D band-stitched magnitude plus 10-D known timing condition. \\
        Window settings & $(L,H)=(10,1)$, $(10,2)$, $(10,5)$, $(10,8)$, $(10,15)$, $(20,5)$. \\
        Split & Chronological 60/20/20; boundary gap $\text{lookback}+\text{horizon}-1$. \\
        Train pools & Measured 6071--6147; synthetic 7156--7182 windows. \\
        Calibration budgets & 50, 100, 300, 600, 1000, 2000, 3000, 4000, 5000 train windows, i.e., about 0.8\%--82\% of the measured pool. \\
        Main model & Slot-conditioned Transformer: 11.4M parameters, hidden 512, 16 heads, 2+2 layers, \gls{ffn} 1024. \\
        Controls & Persistence baseline; shared synthetic-only and shared synthetic-pretrain checkpoints per seed. \\
        Training & Batch 128, 3 seeds; 50 epochs for single-stage training, or 50 synthetic-pretraining epochs followed by 20 measured-data fine-tuning epochs. \\
        \hline
    \end{tabular}
\end{table}

Table~\ref{tab:protocol} summarizes the controlled data-source comparison. The measured training pool contains about 6071--6147 windows depending on the sequence setting, so 50 and 600 windows correspond to about 0.8\% and 10\% of the available measured pool. Within the Transformer evaluation, the three training strategies share the same architecture, evaluation split, and held-out target; only the training data source changes. The pretraining strategy adds offline training, not extra inference capacity. Persistence serves as the no-training reference based on the latest available channel-estimate magnitude. Shared synthetic-only and synthetic-pretrained checkpoints are reused across budgets for each seed and prediction setting.

\subsection{Measurement-Calibrated Synthetic Source}
Here, ``measurement-calibrated'' means that synthetic \gls{csi} is constrained by observable statistics extracted from real \textit{flight-only} measurements. The calibration targets are frame-power level and variability, lag-1 power smoothness, wrapped phase-increment spread, and local temporal and adjacent-subcarrier magnitude consistency. These statistics keep the generator interpretable while grounding it in hardware-inclusive field behavior~\cite{euchner2024ganmimo}.

Synthetic sequences are produced by a generator $g_{\boldsymbol{\phi}}$. Its parameters are selected by matching measured calibration statistics $\mathcal{M}$:
\begin{equation}
    \boldsymbol{\phi}^{*}
    = \arg\min_{\boldsymbol{\phi}}\sum_{m\in\mathcal{M}} w_m
    \frac{|s_m(\mathcal{D}_{S}(\boldsymbol{\phi}))-s_m(\mathcal{D}_{R}^{\mathrm{cal}})|}
    {|s_m(\mathcal{D}_{R}^{\mathrm{cal}})|+\epsilon},
\end{equation}
where $s_m(\cdot)$ denotes a calibration statistic. The final synthetic training set is $\mathcal{D}_{S}=\mathcal{D}_{S}(\boldsymbol{\phi}^{*})$.

The generator is a controllable statistical backbone rather than a black-box generative model. Each synthetic bundle uses 12000 frames over the same 574 retained subcarriers used by the measured representation. The backbone contains four parts: a mission-level power envelope, a smooth frequency profile, a low-frequency temporal-shape field, and a phase random walk. Mathematically, for subcarrier $k$ at time $t$,
\begin{equation}
    \tilde{H}_{t,k}
    = A_t B_k \exp(\alpha S_{t,k})
    \exp\left(j(\Phi_t+\Psi_k)\right),
    \label{eq:synth_backbone}
\end{equation}
where $A_t$ controls the frame-level power envelope, $B_k$ defines the smooth frequency profile, $S_{t,k}$ is a slowly varying temporal-shape field, and $\alpha$ controls its calibrated strength. The random walk $\Phi_t$ captures \gls{lo} drift and residual Doppler, while $\Psi_k$ captures frequency-dependent phase structure. The temporal-shape term is deliberately low frequency in time and subcarrier index: it injects measured-like spectral deformation without giving the synthetic source access to any held-out target frames.

\begin{table}[t]
    \scriptsize
    \centering
    \caption{Synthetic calibration quality.}
    \label{tab:calibration}
    \begin{tabular}{lcc}
        \hline
        Metric & Target & Synthetic \\
        \hline
        Power mean [dB] & 46.723 & 46.723 \\
        Power std [dB] & 4.554 & 4.554 \\
        Power lag-1 corr. & 0.958 & 0.979 \\
        Phase-inc. std [rad] & 1.803 & 1.800 \\
        Temporal similarity mean & 0.9870 & 0.9874 \\
        Temporal similarity std & 0.0258 & 0.0264 \\
        Subcarrier similarity mean & 0.9994 & 0.9987 \\
        Subcarrier similarity std & 0.00075 & 0.00086 \\
        \hline
    \end{tabular}
\end{table}

Table~\ref{tab:calibration} reports calibration against measured calibration-flight statistics; the held-out flight is excluded. Power mean and variance are matched by construction, while phase-increment dispersion and local temporal and adjacent-subcarrier similarity serve as additional consistency checks. Synthetic sequences use zero timing channels where measured data provide slot-id metadata, so the synthetic source cannot access held-out timing labels.

\section{Results and Discussion}
\begin{figure*}[t]
    \centering
    \includegraphics[width=0.96\textwidth]{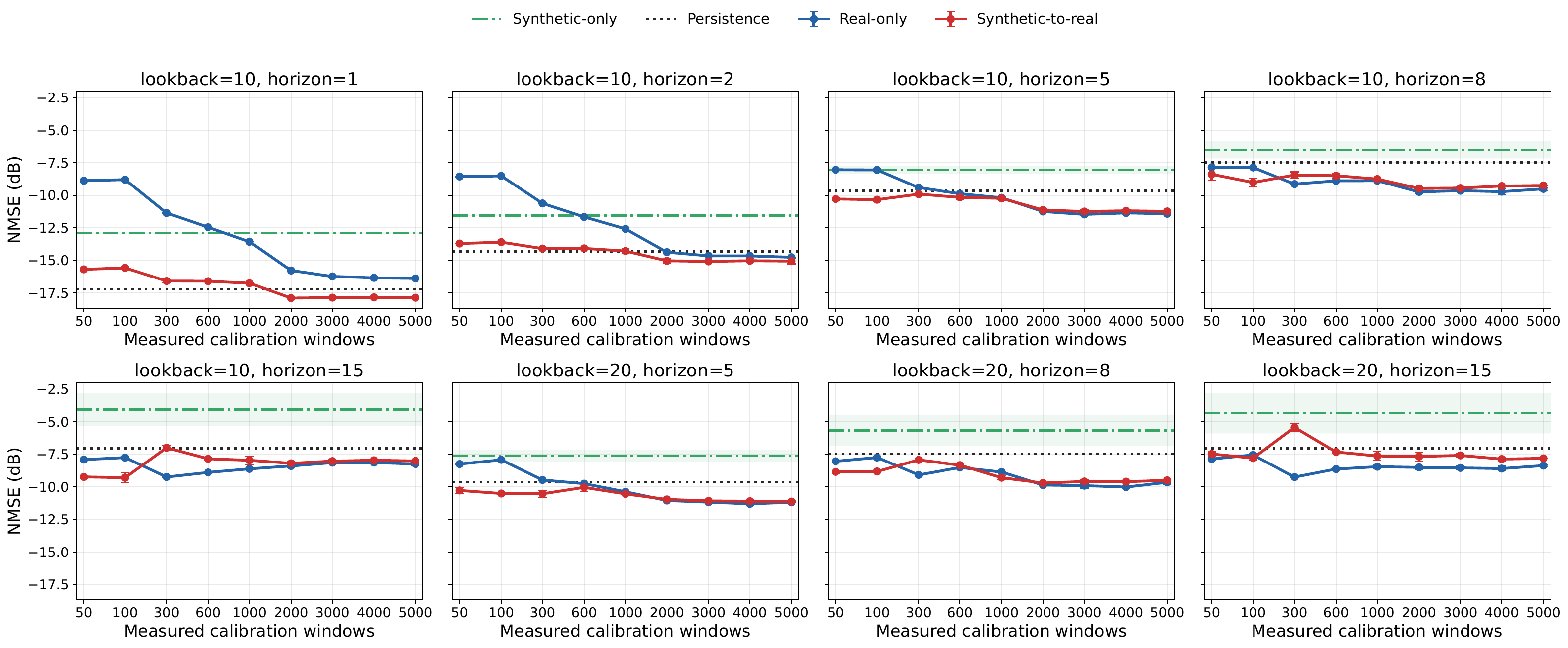}
    \caption{Multiseed \gls{nmse} on the held-out flight under the shared-pretrain protocol. The overview combines the six core settings with \((20,8)\) and \((20,15)\). Error bars show three-seed standard deviations; \textit{synthetic-only} is repeated along the budget axis only for visual comparison.}
    \label{fig:overview}
\end{figure*}

We organize the results around the measured calibration budget and update effort required on a related held-out flight. Fig.~\ref{fig:overview} first uses the Transformer as a high-capacity reference to isolate the data-source effect before target-prefix adaptation. The overview includes the six core protocol settings and two additional long-lookback settings, \((20,8)\) and \((20,15)\), to visualize how the trend extends beyond the aggregate grid. Across the six core settings and nine measured budgets, \textit{\gls{s2r} fine-tuning} most consistently improves the scarce-data regime and crosses persistence more often than \textit{real-only}. The key question is whether this advantage appears before high-data saturation.

The persistence baseline repeats the last observed channel-estimate magnitude vector across the requested horizon. This is a demanding reference for magnitude \gls{csi} because the channel contains a slowly varying component. Table~\ref{tab:breakeven} reports the first measured budget at which each learned strategy crosses it.

\begin{table}[t]
    \scriptsize
    \centering
    \caption{Measured calibration budget required to beat persistence.}
    \label{tab:breakeven}
    \begin{tabular}{lccc}
        \hline
        Setting & Persistence & Real-only & \gls{s2r} \\
        $(L,H)$ & \gls{nmse} [dB] & windows (pool) & windows (pool) \\
        \hline
        $(10,1)$  & -17.22 & $>5000$ ($>81\%$) & 2000 ($\sim$33\%) \\
        $(10,2)$  & -14.33 & 2000 ($\sim$33\%) & 2000 ($\sim$33\%) \\
        $(10,5)$  & -9.66  & 600 ($\sim$10\%) & 50 ($\sim$0.8\%) \\
        $(10,8)$  & -7.48  & 50 ($\sim$0.8\%) & 50 ($\sim$0.8\%) \\
        $(10,15)$ & -7.03  & 50 ($\sim$0.8\%) & 50 ($\sim$0.8\%) \\
        $(20,5)$  & -9.65  & 600 ($\sim$10\%) & 50 ($\sim$0.8\%) \\
        \hline
    \end{tabular}
\end{table}

The break-even budgets show where synthetic pretraining changes the operating point. At horizon 5, i.e., 50\,ms-ahead prediction, \gls{s2r} fine-tuning beats persistence with 50 measured windows, about 0.5\,s and 0.8\% of the measured calibration pool, whereas training from scratch requires 600 windows, about 6\,s and 10\%. For \((10,8)\) and \((10,15)\), both learned strategies already beat persistence at 50 windows, while \((10,1)\) and \((10,2)\) remain harder because persistence is strongest. The result is horizon-dependent calibration acceleration rather than blanket superiority over every short-horizon reference.

\begin{figure}[!t]
    \centering
    \includegraphics[width=\linewidth]{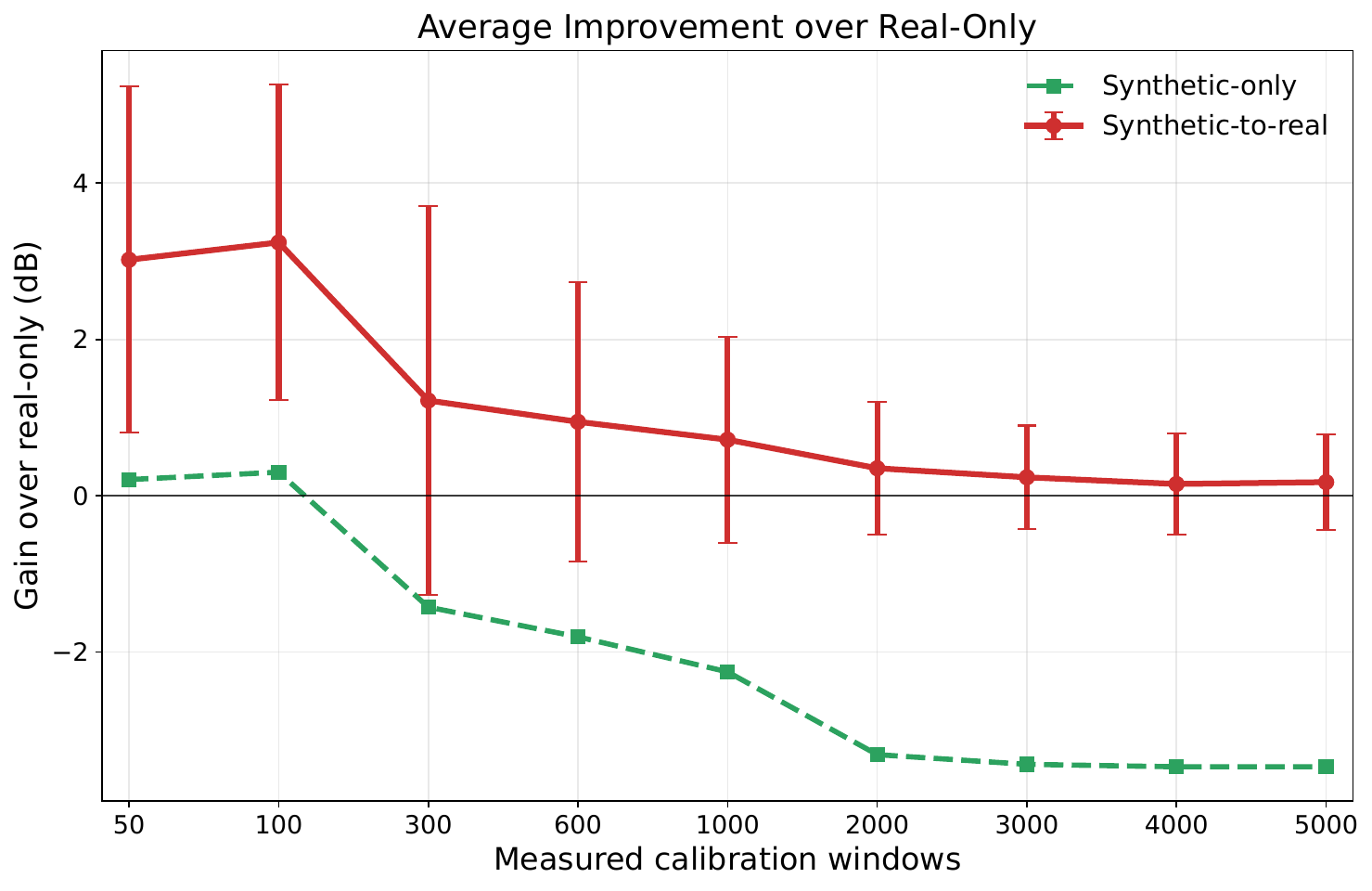}
    \caption{Average improvement over \textit{real-only} versus measured calibration budget over the six core settings.}
    \label{fig:budget_gain}
\end{figure}

Fig.~\ref{fig:budget_gain} shows the same trend averaged across the six core settings. \textit{\gls{s2r} fine-tuning} improves over \textit{real-only} by 3.02\,dB at 50 windows and 3.24\,dB at 100 windows, corresponding to only 0.8\%--1.6\% of the measured calibration pool. The margin then declines to 1.22, 0.95, 0.72, and 0.35\,dB at 300, 600, 1000, and 2000 windows, and remains small at 3000--5000 windows. In contrast, \textit{synthetic-only} does not consistently cross persistence and is therefore useful as a pretraining source, not as a deployable predictor.

The budget curves reveal a saturation regime: once enough measured calibration data are available, the real-only model learns much of the target structure directly and the marginal value of the synthetic prior decreases. The additional \((20,8)\) and \((20,15)\) settings in Fig.~\ref{fig:overview} also show that longer lookback does not by itself guarantee stronger transfer at longer horizons.

\begin{figure}[!t]
    \centering
    \includegraphics[width=\linewidth]{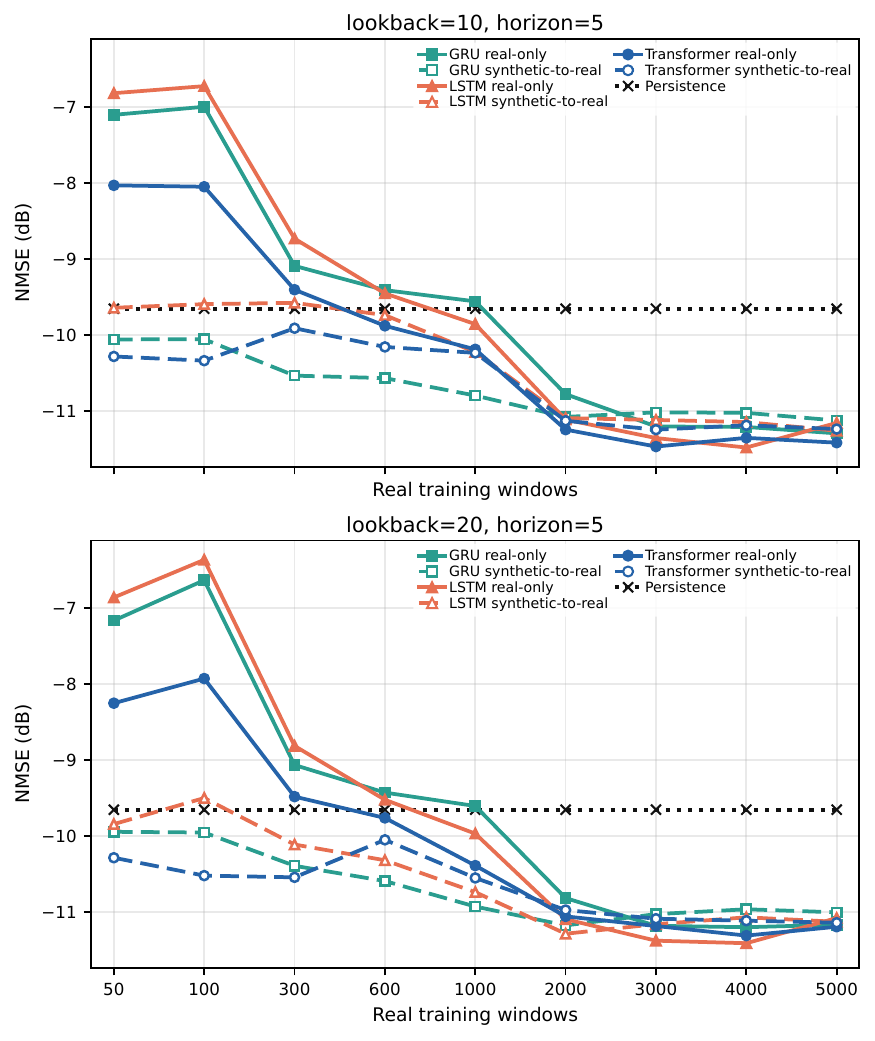}
    \caption{Focused predictor-family control for the two 50\,ms-ahead settings. Color denotes predictor family, line style denotes training strategy, and the dotted black line denotes persistence.}
    \label{fig:family_control}
\end{figure}

A focused predictor-family control then checks whether the effect is tied to the Transformer. Fig.~\ref{fig:family_control} reports the two 50\,ms-ahead settings, \((10,5)\) and \((20,5)\), under the same shared-pretrain protocol, calibration budgets, seeds, and matched training parameters across trainable models. In this control, \gls{gru} has the largest average \gls{s2r} gain over \textit{real-only} at 1.07\,dB, followed by \gls{lstm} at 0.85\,dB and Transformer at 0.58\,dB. The pattern therefore survives beyond a single predictor family.

\begin{figure}[!t]
    \centering
    \includegraphics[width=\linewidth]{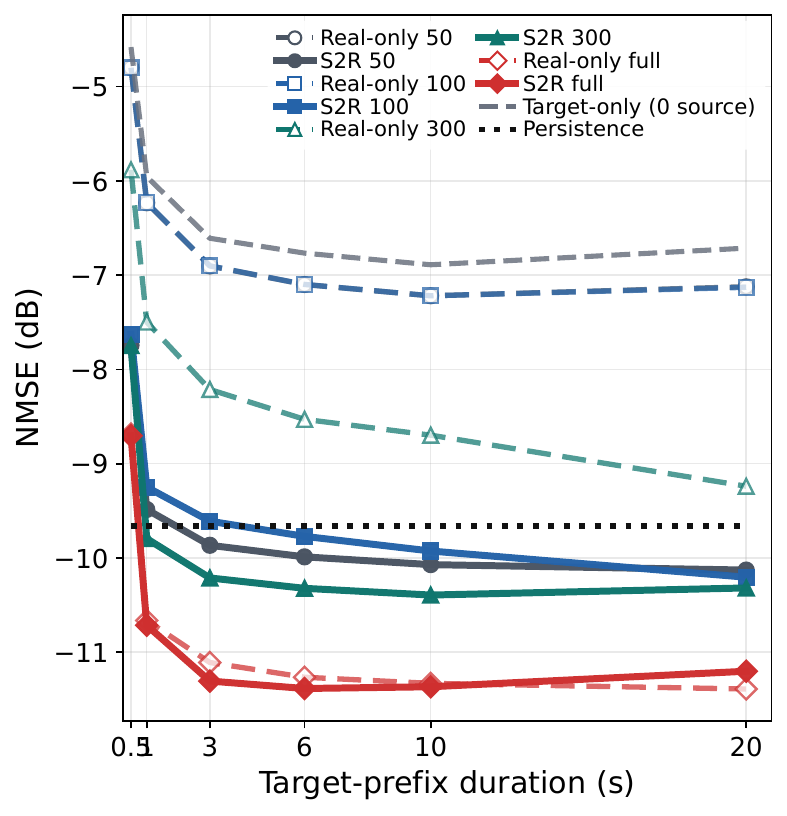}
    \caption{Target-prefix adaptation with \gls{gru} on the held-out flight suffix for 100\,ms of history and 50\,ms-ahead prediction. The legend reports both source-prior route, \textit{real-only} or \textit{\gls{s2r}}, and measured source budget; target-only denotes zero source windows. All learned curves use one prediction-layer update epoch on the target prefix.}
    \label{fig:target_prefix}
\end{figure}

Because \gls{gru} is both lightweight and the strongest focused control in Fig.~\ref{fig:family_control}, Fig.~\ref{fig:target_prefix} uses it for source-prior target-prefix adaptation. The source priors are trained once on the source missions and then reused on a disjoint same-height target flight with different route and speed characteristics. To expose the value of synthetic pretraining when measured source data are also scarce, Fig.~\ref{fig:target_prefix} varies the measured source windows used to build the offline prior. The x-axis varies the target-prefix duration observed after the target flight starts, while each curve label gives the source-prior route and measured source budget. Each learned curve uses one prediction-layer update epoch on the target prefix for the \((10,5)\) setting; the target-only curve corresponds to zero source windows.

Under this more deployment-like protocol, the 100\,ms-history setting shows that a cached prior can be adapted from a short target prefix rather than learned from scratch. When the measured source budget is full, \textit{real-only} and \textit{\gls{s2r}} priors nearly coincide. With only 50--300 measured source windows, however, \gls{s2r} pretraining lowers the target-suffix \gls{nmse} by about 1.7--3.3\,dB across the tested target-prefix durations. Since the target-side update is held fixed, this isolates the quality of the cached source prior; the gain comes from source-side pretraining rather than from a larger deployed network.

Together, the main held-out-flight protocol and the target-prefix check show measurement-calibrated synthetic data acting as a short-calibration prior. In the source-pool evaluation, \textit{real-only} and \textit{\gls{s2r}} use the same \(c\) measured windows; the latter adds only offline pretraining, so inference cost is unchanged within a fixed predictor family. In the target-prefix probe, both source priors are trained before the target flight, and \gls{s2r} is most useful when measured source-flight data are limited rather than abundant.

The present scope is magnitude prediction under related flight conditions. Phase is used only to calibrate synthetic disturbance statistics, and the predictor-family control is focused on 50\,ms-ahead settings rather than a full all-horizon architecture ranking.

\section{Conclusion}
This study shows that measurement-calibrated synthetic pretraining can accelerate adaptation of dynamic \gls{uav} channel predictors. Under a strict held-out protocol, \textit{\gls{s2r} fine-tuning} improves the low-budget regime across the core Transformer grid and crosses persistence earlier than \textit{real-only} in the 50\,ms-ahead settings. A target-prefix check further shows that a cached \gls{s2r} \gls{gru} prior can be adapted with one prediction-layer update epoch, with the clearest advantage when the real-only source prior itself is built from only tens to hundreds of source-flight windows. Once thousands of measured windows are available, training from measured data can approach the pretrained strategy. The strongest use case is therefore scarce source and target adaptation, where synthetic pretraining reduces the measured windows, update effort, and trainable target-side parameters needed to reach a useful predictor.

Future work should make both calibration and adaptation policies more horizon-aware. Two directions follow directly from the present results: extending calibration beyond magnitude-only constraints toward richer complex-domain structure, and designing adaptive policy selection among persistence, real-only learning, and calibrated synthetic pretraining.

\section*{Acknowledgment}
This work of the authors was supported by the German Federal Ministry of Research, Technology, and Space (BMFTR) through the Open6GHub+ project (Grant no. \emph{16KIS2402K}) and the 3D-Net project (Grant no. \emph{16KIS2719}).

\bibliographystyle{IEEEtran}
\bibliography{bibliography}

\end{document}

%% file: acronyms.tex
\newabbreviation[category={common}]{3d}{3D}{dreidimensional}
\newabbreviation[category={common}]{3gpp}{3GPP}{3rd Generation Partnership Project}
\newabbreviation[category={common}]{ai}{AI}{Artificial Intelligence}
\newabbreviation{api}{API}{application programming interface}
\newabbreviation{ar}{AR}{Augmented Reality}
\newabbreviation[category={common}]{cad}{CAD}{computer-aided design}
\newabbreviation{cam}{CAM}{Class Activation Map}
\newabbreviation{drl}{DRL}{deep reinforcement learning}
\newabbreviation{erp}{ERP}{Enterprise Resource Planning}
\newabbreviation{faas}{FaaS}{Function-as-a-Service}
\newabbreviation{fec}{FEC}{forward error correction}
\newabbreviation{ga}{GA}{general aviation}
\newabbreviation[category={common}]{gps}{GPS}{Global Positioning System}
\newabbreviation[category={common}]{gpu}{GPU}{graphics processing unit}
\newabbreviation{h3d}{H3D}{holographic 3D}
\newabbreviation{hmd}{HMD}{head-mounted display}
\newabbreviation{iot}{IoT}{Internet of Things}
\newabbreviation{isv}{ISV}{independent software vendor}
\newabbreviation[category={common}]{it}{IT}{Information Technology}
\newabbreviation{jcas}{JCAS}{Joint Communications and Sensing}
\newabbreviation{kpi}{KPI}{Key Performance Indicator}
\newabbreviation{lidar}{LIDAR}{Light Detection and Ranging}
\newabbreviation{lstm}{LSTM}{long short-term memory}
\newabbreviation{lo}{LO}{local oscillator}
\newabbreviation{mcs}{MCS}{Modulation and Coding Scheme}
\newabbreviation{mr}{MR}{Mixed Reality}
\newabbreviation{ml}{ML}{Machine Learning}
\newabbreviation{moq}{MoQ}{Media over QUIC}
\newabbreviation{nbmp}{NBMP}{network-based media processing}
\newabbreviation[category={common}]{oran}{O-\glsxtrshort{ran}}{Open \glsxtrshort{ran}}
\newabbreviation{poc}{PoC}{proof of concept}
\newabbreviation{pubsub}{Pub/Sub}{Publish/Subscribe}
\newabbreviation{qoe}{QoE}{Quality of Experience}
\newabbreviation{qos}{QoS}{Quality of Service}
\newabbreviation{ran}{RAN}{radio access network}
\newabbreviation[firstplural=radio access technologies]{rat}{RAT}{radio access technology}
\newabbreviation{redcap}{RedCap}{Reduced Capabilities}
\newabbreviation{rem}{REM}{radio environment map}
\newabbreviation[category={common}]{sdk}{SDK}{Software Development Kit}
\newabbreviation{sdr}{SDR}{software-defined radio}
\newabbreviation{slam}{SLAM}{Simultaneous Localization and Mapping}
\newabbreviation{tcas}{TCAS}{traffic alert and collision avoidance system}
\newabbreviation{uav}{UAV}{unmanned aerial vehicle}
\newabbreviation{vr}{VR}{Virtual Reality}
\newabbreviation{xr}{XR}{eXtended Reality}
\newabbreviation{6g}{6G}{sixth-generation}
\newabbreviation{csi}{CSI}{channel state information}
\newabbreviation{usrp}{USRP}{Universal Software Radio Peripheral}
\newabbreviation{oai}{OAI}{Openairinterface5g}
\newabbreviation{nic}{NIC}{network interface card}
\newabbreviation{los}{LoS}{line-of-sight}
\newabbreviation{rssi}{RSSI}{received signal strength indicator}
\newabbreviation{cfr}{CFR}{channel frequency response}
\newabbreviation{srs}{SRS}{sounding reference signal}
\newabbreviation{pusch}{PUSCH}{physical uplink shared channel}
\newabbreviation{5g}{5G}{fifth-generation}
\newabbreviation{ntn}{NTN}{non-terrestrial network}
\newabbreviation{gan}{GAN}{generative adversarial network}
\newabbreviation{gru}{GRU}{gated recurrent unit}
\newabbreviation{s2r}{S2R}{synthetic-to-real}
\newabbreviation{rnn}{RNN}{recurrent neural network}
\newabbreviation{nmse}{NMSE}{normalized mean squared error}
\newabbreviation{ffn}{FFN}{feed-forward network}
\newabbreviation{rmse}{RMSE}{root mean squared error}
\newabbreviation{snr}{SNR}{signal to noise ratio}
\newabbreviation{bler}{BLER}{block error rate}
\newabbreviation{gNB}{gNB}{next Generation NodeB}
\newabbreviation{ue}{UE}{user equipment}
\newabbreviation{cqi}{CQI}{channel quality index}
\newabbreviation{harq}{HARQ}{hybrid automatic repeat request}
\newabbreviation{cpe}{CPE}{common phase error}